\documentclass{amsart}
\usepackage{graphicx}
\usepackage{amscd}
\usepackage{amsmath}
\usepackage{amsfonts}
\usepackage{amssymb}
\usepackage[hang,footnotesize,bf]{caption}

\theoremstyle{plain}
\newtheorem{theorem}{Theorem}

\newtheorem{definition}{Definition}
\newtheorem{example}{Example}

\newtheorem{lemma}{Lemma}

\newtheorem{proposition}{Proposition}
\newtheorem{remark}{Remark}

\numberwithin{equation}{section}

\newcommand{\abs}[1]{\lvert#1\rvert}
\newcommand{\norm}[1]{\lVert#1\rVert}
\def \a{\alpha}

\def \eps{\epsilon}
\def \xte{X_{t}^{(i,\eps_1)}}
\def \xt0{X_{t}^{(i,0)}}

\def \parxei{\frac{\partial}{\partial\eps_1}|_{\eps_1=0}X_t^{(i,\eps_1)}}%PARtial derivative of X^i wrt e
\def \sde{stochastic differential equation }
\def \e{\mathbb{E}}

\def \cov{\mathbb{C}\text{ov}}

\begin{document}
\title[Weak and Strong Taylor methods for numerical solutions of SDEs]
{Weak and Strong Taylor Methods for numerical solutions of stochastic differential equations}
\author{Maria Siopacha and Josef Teichmann}\today

\thanks{Financial support from the Austrian Science Fund (FWF)
under grant P 15889 and the START-prize-grant Y328-N13 is gratefully
acknowledged.
Furthermore this work was financially supported by the Christian
Doppler Research Association (CDG). The authors gratefully
acknowledge a fruitful collaboration and continued support by Bank
Austria and the Austrian Federal Financing Agency ({\"O}BFA) through
CDG}

\address{Department of Mathematical Methods in Economics, Vienna
University of Technology, Wiedner Hauptstrasse 8--10/105--1,
A-1040 Vienna, Austria.}
\email{[josef.teichmann,siopacha]@fam.tuwien.ac.at}

\begin{abstract}
We apply results of Malliavin-Thalmaier-Watanabe for strong and
weak Taylor expansions of solutions of perturbed stochastic
differential equations (SDEs). In particular, we work out weight
expressions for the Taylor coefficients of the expansion. The
results are applied to LIBOR market models in order to deal with
the typical stochastic drift and with stochastic volatility. In
contrast to other accurate methods like numerical schemes for the
full SDE, we obtain easily tractable expressions for accurate
pricing. In particular, we present an easily tractable alternative
to ``freezing the drift'' in LIBOR market models, which has an
accuracy similar to the full numerical scheme. Numerical examples
underline the results.
\end{abstract}
\maketitle

\section{Introduction and Setting}

Let $ (\Omega,\mathcal{F},\mathbb{P}) $ be a probability space
carrying an $N$-dimensional Brownian motion ${(W_t)}_{t \geq 0} $
with a $d \times d$ correlation matrix. We consider smooth curves
$F_{\eps}:\mathbb{R}\to L^2(\Omega;\mathbb{R}^N)$ of random
variables, where $\eps\in\mathbb{R}$ is a parameter. We apply
Taylor theorems to obtain strong approximations of the curve $
F_{\eps} $ at $ \epsilon = 0 $ and we apply partial integration on
Wiener space to obtain weak approximations of the law of $
F_{\epsilon} $ for small values of $ \epsilon $.

We choose the notion \emph{Taylor expansion} instead of asymptotic
expansion in order to point out that the strong method is indeed a
classical Taylor expansion with usual conditions for convergence.
The weak method represents a truncated converging power series in
the parameter $ \epsilon $ if -- for instance -- the payoff
$f:\mathbb{R}^N\to\mathbb{R}$ stems from a real analytic function
and some distributional properties are satisfied.

\section{Weak and strong Taylor methods - Structure Theorems}

We introduce in this section two concepts of approximation.
Consider a curve $ \epsilon \mapsto F_{\epsilon}$, where $
\epsilon \in\mathbb{R}$ and $F_{\eps}\in L^2(\Omega;
\mathbb{R}^N)$.
\begin{definition}\label{def strong taylor}
A \textbf{strong Taylor approximation} of order $n \geq 0$ is a
(truncated) power series
\begin{equation}
\mathbf{T}^n_\epsilon (F_{\eps}):=
\sum_{i=0}^{n}\frac{{\epsilon}^i}{i!} \frac{\partial^i}{\partial
{\epsilon}^i}\Big|_{{\epsilon}=0}F_\epsilon,
\end{equation}
such that
\begin{equation}
\e\big(\abs{F_{\epsilon} -
\mathbf{T}^n_{\epsilon}(F_{\eps})}\big)= o({\epsilon}^n),
\end{equation}
holds true as $ \epsilon \to 0 $.
\end{definition}
\begin{remark}
In our setting a strong Taylor approximation of any order $ n \geq
0 $ of the curve $F_{\eps}$ can always be obtained, see for
instance \cite{krieglmichor}.
\end{remark}
Let $ f:\mathbb{R}^N \to \mathbb{R} $ be a Lipschitz function with Lipschitz constant $K$, then we obtain
\begin{equation}\label{eq lip}
\e \big(\abs{f(F_{\epsilon}) -
f\big(\mathbf{T}^n_{\epsilon}(F_{\eps})\big)}\big)\leq K \e
\big(\norm{F_{\epsilon} -
\mathbf{T}^n_{\epsilon}(F_{\eps})}\big)=K o({\epsilon}^n).
\end{equation}
Equation \eqref{eq lip} does not hold anymore if $ f $ is not
globally Lipschitz continuous. In particular, we observe the
dependence of the right hand side on the Lipschitz constant $ K $.
Hence, truncating an a-priori known Taylor expansion leads to an
error term, which contains the Lipschitz constant and is therefore
not useful for non-Lipschitz claims. The weak method navigates
around this feature by partial integration.
\begin{definition}\label{def weak taylor}
A \textbf{weak Taylor approximation} of order $n \geq 0 $ is a power
series for each bounded, measurable $ f:\mathbb{R}^N \to\mathbb{R}$,
\begin{equation*}
\mathbf{W}^n_{\epsilon}(f,F_{\eps}):=
\sum_{i=0}^{n}\frac{{\epsilon}^i}{i!}\e(f(F_0)\pi_i),
\end{equation*}
where $\pi_i\in L^1(\Omega) $ denote real valued, integrable random variables, such that
\begin{equation*}
\abs{\e\big(f(F_{\epsilon})\big) -
\mathbf{W}^n_{\epsilon}(f,F_{\eps})} = o({\epsilon}^n).
\end{equation*}
\end{definition}
\begin{remark}
The weights $ \pi_i $ for $ i \geq 1 $ are called \emph{Malliavin weights}.
\end{remark}
\begin{remark}
If the law of $ F_{\epsilon} $ is real analytic at $ \epsilon = 0
$ in the weak sense, i.e.~ if there exist (signed) measures $
\mu_i $ such that for all bounded, measurable $ f:\mathbb{R}^N \to
\mathbb{R} $ the following series converges and the equality
\begin{equation*}
\e\big(f(F_{\epsilon})\big) = \sum_{i\geq 0}
\frac{{\epsilon}^i}{i!} \int_{\mathbb{R}^N} f(x) \mu_i(dx),
\end{equation*}
holds true, precisely then we do have a converging weak Taylor
expansion. We aim for constructing stochastic representations of
the following type, for $i\geq 0$:
\begin{equation*}
\int_{\mathbb{R}^N} f(x) \mu_i (dx) = \e(f(F_0) \pi_i).
\end{equation*}
\end{remark}

For the definition of the weak Taylor approximation to make sense,
existence of the Malliavin weights has to hold. The following
theorem can be found in a slightly different version in
\cite{malliavinthalmaier} and goes back to S.~Watanabe. For the
definition and notion of $\mathcal{D}^{\infty}(\mathbb{R}^N)$ see
\cite{malliavin} or \cite{nualartbook}.
\begin{theorem}
Let $F_{\eps}:\mathbb{R}\to\mathcal{D}^{\infty}(\mathbb{R}^N)$ be
smooth and assume that the Malliavin covariance matrix
$\gamma(F_{\eps})$ is invertible with $p$-integrable inverse for
every $p \geq 1$ around $\epsilon=0$ (i.e.~on an open interval
containing $ \epsilon = 0 $). Then there is a weak Taylor
approximation of any order $ n \geq 0 $ and there are explicit
formulas for the weights $ \pi_i $. If we only know that the
Malliavin covariance matrix $\gamma(F_0)$ is invertible with
$p$-integrable inverse, then we can also calculate the Malliavin
weights, since they depend only on $\gamma(F_0)$.
\end{theorem}
\begin{proof}
Fix $ n \geq 0 $ and take a smooth test function $ f:\mathbb{R}^N
\to \mathbb{R} $ and assume that $\gamma^{-1} (F_{\epsilon})$ exists
as a smooth curve in $ \mathcal{D}^{\infty} $ on a open
$\epsilon$-interval containing $ \epsilon = 0 $. By standard
arguments we can prove the following formula
\begin{equation*}
\frac{d}{d \epsilon} \e\big(f(F_{\epsilon})\big)= \e\Bigg(
f(F_{\epsilon}) \delta \Big( s \mapsto {(D_s
F_{\epsilon})}^{\mathrm{T}} \gamma^{-1} (F_{\epsilon})\frac{d}{d
\epsilon} F_{\epsilon}\Big) \Bigg).
\end{equation*}
More precisely, by the integration by parts \cite[Definition
1.3.1-(1.42)]{nualartbook}, the chain rule \cite[Proposition
1.2.3]{nualartbook} and the definition of the Malliavin covariance
matrix, \cite[page 92]{nualartbook}, we obtain from the right hand
side the desired left hand side.

Notice that the $\epsilon$-dependence of the Skorohod integral is
smooth due to basic properties of $ \mathcal{D}^{\infty} $. Hence,
we can calculate higher derivatives of the left hand side by
iterating the above procedure and differentiating the Skorohod
integral. We denote
\begin{equation}\label{eq pi1}
\pi_1 :=\delta \Big( s \mapsto {(D_s F_{\epsilon})}^{\mathrm{T}}
\gamma^{-1}(F_{\epsilon})\frac{d}{d \epsilon} F_{\epsilon}\Big).
\end{equation}
We write then, pars pro toto, the formula for the second
derivative
\begin{align*}
\frac{d^2}{d {\epsilon}^2} \e(f(F_{\epsilon})) & =  \e\Bigg(
f(F_{\epsilon}) \delta \Big( s \mapsto \pi_1{(D_s
F_{\epsilon})}^{\mathrm{T}} \gamma^{-1} (F_{\epsilon})
\frac{d F_{\epsilon}}{d \epsilon} \Big) \Bigg)+ \\
& + \e\Bigg( f(F_{\epsilon}) \delta \Big( s \mapsto {(D_s
\frac{dF_{\epsilon}}{d \epsilon})}^{\mathrm{T}}
\gamma^{-1} (F_{\epsilon}) \frac{d F_{\epsilon}}{d \epsilon} \Big) \Bigg) - \\
& - \e\Bigg( f(F_{\epsilon}) \delta\Big(s \mapsto {(D_s
F_{\epsilon})}^{\mathrm{T}} \gamma^{-1} (F_{\epsilon}) \frac{d
\gamma(F_{\epsilon})}{d \epsilon}\gamma^{-1} (F_{\epsilon})
\frac{d F_{\epsilon}}{d \epsilon} \Big) \Bigg)+ \\
&  + \e\Bigg( f(F_{\epsilon}) \delta\Big(s \mapsto {(D_s
F_{\epsilon})}^{\mathrm{T}} \gamma^{-1} (F_{\epsilon}) \frac{d^2
F_{\epsilon}}{d {\epsilon}^2} \Big) \Bigg).
\end{align*}
This formula makes perfect sense at $ \epsilon = 0 $ and -- by
induction -- we see that we can perform this step for any
derivative. The general, recursive result is the following:
\begin{align*}
a_s & := {(D_s F_{\epsilon})}^{\mathrm{T}} \gamma^{-1}
(F_{\epsilon})
\frac{d F_{\epsilon}}{d \epsilon} \text{ for } 0 \leq s \leq T, \\
\pi_n & := \delta( s \mapsto a_s \pi_{n-1} ) + \frac{d}{d \epsilon} \pi_{n-1}, \\
\pi_0 & := 1.
\end{align*}
Here we understand the weights $ \pi_n $ as $\epsilon$-dependent,
whereas in the final formulas we put $\epsilon =0$. This proves
the result for smooth test functions $ f $ and under the
assumption that the Malliavin covariance matrix is invertible
around $ \epsilon = 0 $. If we approximate a bounded, measurable
function $ f $ by smooth test functions we obtain the desired
assertion by standard arguments, since the weights are integrable.
\end{proof}
\begin{remark}
By Taylor's theorem and the Fa\`{a}-di-Bruno-formula we obtain
$$
\frac{d^n f(F_{\epsilon})}{d {\epsilon}^n} = \sum_{|\alpha|\leq n}
f^{(\alpha)}(F_{\epsilon}) p_{\alpha},
$$
where $ p_{\alpha} $ is a well-defined polynomial in derivatives
of the curve $ \epsilon \mapsto F_{\epsilon}$, for a multi-index
$\a$. Since $ \mathcal{D}^{\infty} $ is an algebra, see
\cite{malliavin}, the above expression lies in $ L^p(\Omega) $ for
each $ p \geq0 $. The previous result provides a representation of
the partial integration result for
$$
\e(\sum_{|\alpha|\leq n} f^{(\alpha)}(F_{\epsilon}) p_{\alpha})=
E(f(F_{\epsilon}) \pi_n ).
$$
The structure of the weights is seen from above. The result can be
considered as a dual version of the Fa\`{a}-di-Bruno-formula. However,
the structure of this dual formula is much simpler.
\end{remark}

We provide an example to demonstrate the strong and weak method of
approximation. The method works in order to replace time-consuming
iteration schemes, like the Euler-scheme, by simulations of
``simple'' It\^{o} integrals.
\begin{example}
We deal with a generic, real-valued random variable over a
one-dimensional Gaussian space, see \cite{nualartbook}, i.e.
$$
F_{\epsilon} = \sum_{i=0}^{\infty} \frac{{\epsilon}^i}{i!}F^{i},
$$
where the $F^{i}$ lie in the $(i+1)^{st}$ Wiener chaos
$\mathcal{H}_{i+1}(\Omega)$ (one can think of a Hermite expansion
for instance) and the sum is understood in the $ L^2 $-sense. From
the strong expansion we obtain immediately -- for a given
Lipschitz function $ f :\mathbb{R} \to \mathbb{R} $ -- that
$$
\abs{\e\big(f(F_{\epsilon})\big) - \e\big(f(F^{0} + \epsilon
F^{1})\big)} \leq K o(\epsilon),
$$
as $\epsilon\to 0$, where $K$ denotes the Lipschitz constant of $
f $. This simple approximation can be sometimes quite useful.

We assume now that $ F^{0} = \int_0^{\infty} h(s) dW_s $ has
non-vanishing variance in order to calculate the weights, which do
depend only on $ \gamma(F_0) $. The strong Taylor approximation is
given by definition, the weak Taylor expansion can be constructed
by the previous recursive formulas and the specifications
\begin{align*}
&D_s F^{0} = h(s), \\
&\gamma(F^{0}) = \int_0^{\infty} h(s)^2 ds, \\
&a_s = \frac{h(s)}{\int_0^{\infty} h(s)^2 ds}.
\end{align*}
In order to obtain a first-order approximation for bounded,
measurable random variables we therefore have to calculate
$$
\e\big(f(F^{0})\big) + \epsilon \e\big(f(F^{0})\pi_1\big),
$$
where
$$
\pi_1 = \delta \big(s \mapsto a_s F^{1}\big).
$$
This amounts to an integration of $f$ times a polynomial with
respect to a Gaussian density, since:
$$
\e\big(f(F^{0}) \pi_1\big) = \e\big(f(F^{0}) F^{1} \int_0^{\infty}
a_s dW_s\big) - \int_0^{\infty}\e\big(f(F^{0}) D_s F^{1}
a_s\big)ds.
$$
Notice that the strong approximation does not yield such a result
for \emph{bounded, measurable random variables}. Notice also that
in the given case the approximation can be calculated in a
deterministic way, since we deal with Gaussian integrations.

The second-order weak Taylor approximation is given by
\begin{equation*}
\e\big(f(F^{0})\big)+\epsilon\e\big(f(F^{0})\pi_1\big)+\epsilon^2\e\big(f(F^{0})\pi_2\big),
\end{equation*}
where
$$
\pi_2 = \delta \big(s \mapsto \pi_1 a_s F^{1}\big)+ \delta \big(s
\mapsto a_s F^{2}\big).
$$
\end{example}
%%%%%%%%%%%%%%%%%%%%%%%%%%%%%%%%%%%%%%%%%%%%%%%%%%%%%%%%%%%%%%%%%%%%%%%%%%%%%%%%%%%%%%%%%%%%%%%
\section{Applications from Financial Mathematics}
For applications we want to deal with strong and weak Taylor
approximations of a given curve of random variables. We are
particulary interested in cases, where the first derivative $
\frac{d F_{\epsilon}}{d \epsilon}|_{\epsilon = 0} $ is of simple
form or -- even more important -- where the Malliavin covariance
matrix $ \gamma(F_0) $ is of simple form. In these cases it is
easy to obtain first or second order approximations of the
respective quantities in the weak or strong sense.

In what follows, first we will present one of the most applied
interest rate models, namely the LIBOR market model (LMM). Then,
we will introduce the commonly used technique of \emph{freezing
the drift}. We will show how to embed the''freezing the drift"
technique into our framework of Taylor approximations. We
understand freezing the drift as a strong Taylor approximation of
order zero in the drift term of the LIBOR SDE. Our goal is to put
this technique into a method, where we can in particular improve
the order of approximation. We will finally extend the assumption
of log normality and develop a stochastic volatility LMM, where we
will show how to obtain tractable option prices via our weak
Taylor approximations.

\subsection{The LIBOR Market Model} We apply our concepts to the
LMM, initially constructed by \cite{bgm97}, \cite{mss97} and
\cite{jamshidian97}. Let $T$ denote a strictly positive fixed time
horizon and $(\Omega, \mathcal{F}_T, \mathbb{P},
(\mathcal{F}_{t})_{0\leq t\leq T})$ be a complete probability
space, supporting an $N$-dimensional Brownian motion
$W_t=(W_t^{1},...W_t^{N})_{0\leq t\leq T}$. The factors are
correlated with $dW_t^{i}$ $dW_t^{j}=\rho_{ij}dt$. Let $0=T_0 <
T_{1} < T_{2} < \ldots < T_{N} < T_{N+1}=:T$ be a discrete tenor
structure and $\a:=T_{i+1}-T_i$ the accrual factor for the time
period $[T_{i}, T_{i+1}]$, $i=0,\ldots, N$. Let $P(t,T_i)$ denote
the value at time $t$ of a zero coupon bond with maturity
$T_i\in[0,T]$. The measure $\mathbb{P}$ is the terminal forward
measure, which corresponds to taking the final bond $P(t,T)$ as
num\'{e}raire. The forward LIBOR rate $L_t^i:=L_t(T_i,T_{i+1})$ at
time $t\leq T_i$ for the period $[T_{i},T_{i+1}]$ is given by:
\begin{equation*}
L_{t}^{i}=L_{t}(T_{i},T_{i+1})=\frac{1}{\a}\bigg(
\frac{P(t,T_{i})}{P(t,T_{i+1})}-1\bigg).
\end{equation*}
We assume that for any maturity $T_{i}$ there exists a bounded,
continuous, deterministic function
$\sigma^i(t):[0,T_{i}]\to\mathbb{R}$, which represents the
volatility of the LIBOR $L_{t}^{i}$, $i=1,...,N$. The log normal
LIBOR market model can be expressed under the measure $\mathbb{P}$
as:
\begin{equation}\label{standard LMM}
dL_{t}^{i} = \sigma^{i}(t)L_{t}^{i}\Big(-\sum_{j=i+1}^{N}\frac{\a
L_{t}^{j}\sigma^{j}(t)} {1+\a
L_{t}^{j}}\rho_{ij}\Big)dt+\sigma^{i}(t)L_{t}^{i}dW_{t}^{i},i=1,...,N.
\end{equation}

\subsection{Freezing the Drift} The dynamics of forward LIBORs for
$i=1,...,N-1$ depend on the stochastic drift term $\frac{\a
L_{t}^{j}}{1+\a L_{t}^{j}}$, $i \leq j \leq N$, which is
determined by LIBOR rates with longer maturities. This random
drift prohibits analytic tractability when pricing products that
depend on more that one LIBOR rate, since there is no unifying
measure under which all LIBOR rates are simultaneously log normal.
In addition, it encumbers the numerical implementation of the
model. Common practice is to approximate this term by its starting
value $\frac{\a L_{0}^{j}}{1+\a L_{0}^{j}}$ or as it is widely
referred to as \emph{freezing the drift}, i.e.
\begin{equation*}
\frac{\a L_{t}^{j}}{1+\a L_{t}^{j}} \approx \frac{\a
L_{0}^{j}}{1+\a L_{0}^{j}}.
\end{equation*}
It was first implemented in the original paper \cite{bgm97} for
the pricing of swaptions based on the LMM. \cite{bracewomer} and
\cite{schlogl02} argue that freezing the drift is justified due to
the fact that this term has small variance. However, by freezing
the drift there is a difference in option prices with the real and
the frozen drift. It has not been examined how big the error is or
for which assets it works well or not. Our aim is to investigate
such a phenomenon and improve the performance by providing with
correction terms of order one.
%%%%%%%%%%%%%%%%%%%%%%%%%%%%%%%%%%%%%%%%%%%%%%%%%%%%%%%%%%%%%%%%%%%%%%%%%%%%%%%%%
\subsection{Correcting the Frozen Drift}\label{section freeze
drift} The purpose of this section is to embed the well-known and
often applied technique of \emph{freezing the drift} into the
strong and weak Taylor approximations, in order to develop a
method to improve the order of accuracy. Specifically for the
strong Taylor approximation, the method works well, since we
always deal with a globally Lipschitz drift term $x\mapsto\frac{\a
x_{+}}{1+\a x_{+}}$ with small Lipschitz constant $\a$.
\begin{remark}
As it will be clear later, the strong Taylor correction method can
be accommodated with any extension of the log normal LMM, for
example with the L\'{e}vy LIBOR model by Eberlein and {\"O}zkan
\cite{eberoz}.
\end{remark}
%%%%%%%%%%%%%%%%%%%%%%%%%%%%%%%%%%%%%%%%%%%%%%%%%%%%%%%%%%%%%%%%%%%%%%%%%%%%%%%%%%%%%%%%%%%%%
\subsubsection{Strong Taylor Approximation}
We first state a useful
lemma, asserting that we can indeed freeze the drift under special
model formulation and choice parameters.
\begin{lemma}\label{lemma freeze drift}
Let $\eps_1\in\mathbb{R}$ and consider for $i=1, \ldots,N$ the
following stochastic differential equation:
\begin{align}\label{eq sde e-model}
dX_{t}^{(i,\eps_1)}&=\eps_1\Bigg(
\sigma^{i}(t)X_{t}^{(i,\eps_1)}\Big(-\sum_{j=i+1}^{N} {\frac{\a
X_{t}^{(j,\eps_1)}\sigma^{j}(t)} {1+\a X_{t}^
{(j,\eps_1)}}}\rho_{ij}dt+dW_{t}^{i}\Big)\Bigg),
\end{align}
defined on the complete probability space $(\Omega, \mathcal{F}_T,
\mathbb{P}, (\mathcal{F}_{t})_{0\leq t\leq T})$ where $W_{t}$ is
an $N$-dimensional Brownian motion under the measure $\mathbb{P}$
with $dW_t^idW_t^j=\rho_{ij}dt$. Then the first-order strong
Taylor approximation for $X_t^{(i,\eps_1)}$ is given by:
\begin{equation}\label{T1}
\mathbf{T}_{\eps_1}^1(X_t^{(i,\eps_1)})=\xt0+\eps_1\frac{\partial}{\partial\eps_1}\Big|_{\eps_1=0}X_t^{(i,\eps_1)}.
\end{equation}
\begin{proof}
By \eqref{def strong taylor} we obtain for $n=1$:
\begin{equation*}
\mathbf{T}_{\eps_1}^1(X_t^{(i,\eps_1)})\simeq\xte=X_{0}^{(i,0)}+\eps_1Y^i_t+o(\eps_1),
\end{equation*}
since $X_{t}^{(i,0)}=X_{0}^{(i,0)}$ and where $Y^i_t:=\parxei$ is
the first-order correction term. By differentiating \eqref{eq sde
e-model} with respect to $\eps_1$, we calculate:
\begin{align*}
d\Big(\frac{\partial}{\partial\eps_1}\Big|_{\eps_1=0}X_t^{(i,\eps_1)}\Big)
&=\sigma^{i}(t)X_{0}^{(i,0)}\Big(-\sum_{j=i+1}^{N}\frac{\a
X_{0}^{(j,0)}\sigma^{j}(t)}{1+\a
X_{0}^{(j,0)}}\rho_{ij}dt+dW_{t}^{i}\Big),
\end{align*}
and derive $Y^i_t$ as the solution to the above linear SDE:
\begin{equation}\label{Y_t^i}
Y^i_t=\int_{0}^{t}-\sigma^{i}(s)X_{0}^{(i,0)}\Big(\sum_{j=i+1}^{N}\frac{\a
X_{0}^{(j,0)}\sigma^{j}(s)}{1+\a X_{0}^{(j,0)}}\rho_{ij}\Big)ds+
\int_{0}^{t}\sigma^{i}(s)X_{0}^{(i,0)}dW_{s}^{i},
\end{equation}
with $ Y^i_0=0$.
\end{proof}\end{lemma}

\begin{remark}
We parametrise the LIBOR market model in terms of the parameter
$\eps_1$ as follows:
\begin{equation*}
dL_{t}^{(i,\eps_1)}=
\sigma^{i}(t)L_{t}^{(i,\eps_1)}\Big(-\sum_{j=i+1}^{N} {\frac{\a
X_{t}^{(j,\eps_1)}\sigma^{j}(t)}{1+\a
X_{t}^{(j,\eps_1)}}}\rho_{ij}dt+dW_{t}^{i}\Big).
\end{equation*}
and assume at $t=0$ that $L_0^{(i,\eps_1)}=X_0^{(i,\eps_1)}$ for
all $\eps_1$ and all $i=1,...,N$. If $\eps_1=1$, what we obtain is
the standard LIBOR market model formulation and in particular
$L_{t}^{(i,1)}= X_{t}^{(i,1)}$. For $\eps_1=0$, $\xt0$ equals its
starting value and thus the drift term in the following SDE is no
longer stochastic:
\begin{equation}\label{frozen sde}
dL_{t}^{(i,0)}= \sigma^{i}(t)L_{t}^{(i,0)}\Big(-\sum_{j=i+1}^{N}
{\frac{\a X_{t}^{(j,0)}\sigma^{j}(t)}{1+\a
X_{t}^{(j,0)}}}\rho_{ij}dt+dW_{t}^{i}\Big).
\end{equation}
\end{remark}

The next proposition provides a way for a pathwise approximation
of $L_t^{(i,\eps_1)}$, by means of adjusting its SDE. This is
achieved by adding $\mathbf{T}_{\eps_1}^n(X_t^{(j,\eps_1)})$ in
the frozen drift part.
\begin{proposition}\label{prop strong drift}
Assume the setup of Lemma \ref{lemma freeze drift} and assume
further at $t=0$ that $L_0^{(i,\eps_1)}=X_0^{(i,\eps_1)}$ for all
$\eps_1$ and all $i=1,...,N$. Then the \sde for
$L_{t}^{(i,\eps_1)}$ with the unfrozen drift:
\begin{equation}\label{eq sde without strong proxy}
dL_{t}^{(i,\eps_1)}=
\sigma^{i}(t)L_{t}^{(i,\eps_1)}\Big(-\sum_{j=i+1}^{N} {\frac{\a
X_{t}^{(j,\eps_1)}\sigma^{j}(t)} {1+\a
X_{t}^{(j,\eps_1)}}}\rho_{ij}dt+dW_{t}^{i}\Big),
\end{equation}
can be strongly approximated as $\eps_1\downarrow 0$ by
\begin{equation}\label{eq sde strong proxy}
d\hat{L}_{t}^{(i,\eps_1)}=
\sigma^{i}(t)\hat{L}_{t}^{(i,\eps_1)}\Big(-\sum_{j=i+1}^{N}\frac{\a
\big(\mathbf{T}_{\eps_1}^n(X_t^{(j,\eps_1)})\big)_{+}\sigma^{j}(t)}{1+\a
\big(\mathbf{T}_{\eps_1}^n(X_t^{(j,\eps_1)})\big)_{+}}\rho_{ij}dt+dW_{t}^{i}\Big).
\end{equation}
\begin{remark} For $n=0$, we derive the ''freezing the drift" case. For
$n=1$, we already obtain an improvement.
\end{remark}
\begin{proof}
First step is to interchange $X_{t}^{(j,\eps_1)}$ with
$(X_{t}^{(j,\eps_1)})_{+}$ in \eqref{eq sde without strong proxy}
to obtain:
\begin{equation*}
dL_{t}^{(i,\eps_1)}=
\sigma^{i}(t)L_{t}^{(i,\eps_1)}\Big(-\sum_{j=i+1}^{N} {\frac{\a
(X_{t}^{(j,\eps_1)})_{+}\sigma^{j}(t)} {1+\a
(X_{t}^{(j,\eps_1)})_{+}}}\rho_{ij}dt+dW_{t}^{i}\Big).
\end{equation*}
This yields no change for the dynamics of $L_{t}^{(i,\eps_1)}$,
since $X_{t}^{(j,\eps_1)}=(X_{t}^{(j,\eps_1)})_{+}$.

By Taylor's expansion, we know that as $\eps_1\downarrow 0$,
$\hat{L}_{t}^{(i,\eps_1)}\rightarrow L_{t}^{(i,\eps_1)}$
$\mathbb{P}$-a.s. The estimate for the error term is given by
\begin{align*}
&\log{\hat{L}_{t}^{(i,\eps_1)}}-\log{L_{t}^{(i,\eps_1)}}=\\&=\int_{0}^{t}
\sigma^{i}(s)\Big(-\sum_{j=i+1}^{N}\frac{\a
\big(\mathbf{T}_{\eps_1}^n(X_s^{(j,\eps_1)})\big)_{+}\sigma^{j}(s)}{1+\a
\big(\mathbf{T}_{\eps_1}^n(X_s^{(j,\eps_1)})\big)_{+}}\rho_{ij}+\sum_{j=i+1}^{N}
{\frac{\a (X_{t}^{(j,\eps_1)})_{+}\sigma^{j}(t)} {1+\a
(X_{t}^{(j,\eps_1)})_{+}}}\rho_{ij}\Big)ds\leq \\
&\leq
\int_{0}^{t}\a\abs{X_s^{(j,\eps_1)}-(\mathbf{T}_{\eps_1}^n(X_s^{(j,\eps_1)}))_+}ds.
\end{align*}
\end{proof}\end{proposition}

\begin{remark} The SDE for the approximated $\hat{L}_{t}^{(i,\eps_1)}$
is easier and faster to simulate than (\ref{standard LMM}), as it
is exhibited by the following example. Notice additionally that
$\hat{L}_{t}^{(i,\eps_1)}$ is a continuous functional of the
process $Y^j_t$ \eqref{Y_t^i} and of the Brownian path $W_t^i$.
Eventually, by using $\hat{L}_{t}^{(i,\eps_1)}$ as the LIBOR
rates, the computational complexity of the drift and thus of the
model can be reduced substantially, while maintaining accuracy of
prices.
\end{remark}

\begin{example} In this example, we examine the performance of the
strong Taylor correction method. Let $N=3$ and consider pricing a
caplet on the LIBOR rate $L^1$ with strike $K$. Its price is given
by:
\begin{equation*}
P^{cpt}_0=\a \e_{\mathbb{P}}\Big(\big(L_{T_1}^{1}-K\big)_+\Big).
\end{equation*}
Assume that the volatility functions
$\sigma^i(t):[0,T_{i}]\rightarrow\mathbb{R}$ for $i=1,2,3$ are
given by (cf. Brigo and Mercurio \cite{brigomercurio}, formulation
(6.12)):
\begin{equation*}
\sigma^i(t)=\big(a(T_i-t)+d\big)\exp{\big(-b(T_i-t)\big)}+e,
\end{equation*}
where the constants $a,b,d,e$ are the same for all three LIBOR
rates and are equal to $a=-0.113035$, $b=0.22911$, $d=-a$,
$e=0.684784$. Thus, we can write the model under the terminal
measure $\mathbb{P}$ as:
\begin{align*}
dL_{t}^{(1,\eps_1)}&=\sigma^1(t)L_{t}^{(1,\eps_1)}\Big(-{\frac{\a
X_{t}^{(2,\eps_1)}\sigma^2(t)\rho_{12}} {1+\a
X_{t}^{(2,\eps_1)}}}-{\frac{\a
X_{t}^{(3,\eps_1)}\sigma^3(t)\rho_{13}} {1+\a
X_{t}^{(3,\eps_1)}}}\Big)dt+\sigma^1(t)L_{t}^{(1,\eps_1)}dW_{t}^{1},\\
dL_{t}^{(2,\eps_1)}&=\sigma^2(t)L_{t}^{(2,\eps_1)}\Big(-{\frac{\a
X_{t}^{(3,\eps_1)}\sigma^3(t)\rho_{23}} {1+\a
X_{t}^{(3,\eps_1)}}}\Big)dt+\sigma^2(t)L_{t}^{(2,\eps_1)}dW_{t}^{2},\\
dL_{t}^{3}&=\sigma^3(t)L_{t}^{3}dW_{t}^{3},\\
dX_{t}^{(1,\eps_1)}&=\eps_1\Big(\sigma^1(t)
X_{t}^{(1,\eps_1)}\Big(-{\frac{\a
X_{t}^{(2,\eps_1)}\sigma^2(t)\rho_{12}} {1+\a
X_{t}^{(2,\eps_1)}}}-{\frac{\a
X_{t}^{(3,\eps_1)}\sigma^3(t)\rho_{13}} {1+\a
X_{t}^{(3,\eps_1)}}}\Big)dt+\sigma^1(t)X_{t}^{(1,\eps_1)}dW_{t}^{1}\Big),\\
dX_{t}^{(2,\eps_1)}&=\eps_1\Big(\sigma^2(t)
X_{t}^{(2,\eps_1)}\Big(-{\frac{\a
X_{t}^{(3,\eps_1)}\sigma^3(t)\rho_{23}} {1+\a
X_{t}^{(3,\eps_1)}}}\Big)dt+\sigma^2(t)X_{t}^{(2,\eps_1)}dW_{t}^{2}\Big),\\
dX_{t}^{(3,\eps_1)}&=\eps_1\Big(\sigma^3(t)
X_{t}^{(3,\eps_1)}dW_{t}^{3}\Big),
\end{align*}
with initial values $L_0^{(i,\eps_1)}=X_0^{(i,\eps_1)}=c_i$, for
$i=1,2,3$ and for all $\eps_1$. The Brownian motion vector
$(W_{t}^{1},W_{t}^{2},W_{t}^{3})$ is correlated with correlation
coefficient $\rho_{ij}$ given by:
\begin{equation*}
\rho_{ij}=0.49+(1-0.49)\exp{(-0.13\abs{i-j})}, \ i,j=1,2,3.
\end{equation*}
The SDEs for the approximated LIBOR rates
$\hat{L}_{t}^{(1,\eps_1)}$ and $\hat{L}_{t}^{(2,\eps_1)}$ are
given by:
\begin{align*}
d\hat{L}_{t}^{(1,\eps_1)}&=\sigma^1(t)\hat{L}_{t}^{(1,\eps_1)}\Big(-\frac{\a
\big(c_2+\eps_1Y_t^2\big)_+\sigma^2(t)\rho_{12}}{1+\a
\big(c_2+\eps_1Y_t^2\big)_+}-\frac{\a\big(c_3+\eps_1Y_t^3\big)_+\sigma^3(t)\rho_{13}}
{1+\a\big(c_3+\eps_1Y_t^3\big)_+}\Big)dt+\\
&+\sigma^1(t)\hat{L}_{t}^{(1,\eps_1)}dW_{t}^{1},\\
d\hat{L}_{t}^{(2,\eps_1)}&=\sigma^2(t)\hat{L}_{t}^{(2,\eps_1)}\Big(-\frac{\a
\big(c_3+\eps_1Y_t^3\big)_+\sigma^3(t)\rho_{23}} {1+\a
\big(c_3+\eps_1Y_t^3\big)_+}\Big)dt+\sigma^2(t)\hat{L}_{t}^{(2,\eps_1)}dW_{t}^{2}.
\end{align*}
The partial derivative terms $Y_t^2$ and $Y_t^3$ are equal to:
\begin{align*}
Y_t^2&=c_2\Big(\int_{0}^{t}\sigma^{2}(s)dW_{s}^{2} -\frac{\a
c_3\rho_{23}}{1+\a
c_3}\int_{0}^{t}\sigma^{2}(s)\sigma^{3}(s)ds\Big),\\
Y_t^3&=c_3\int_{0}^{t}\sigma^{3}(s)dW_{s}^{3}.
\end{align*}
We compare three caplet prices:
\begin{itemize}
\item benchmark price, underlying $L_{t}^{(1,\eps_1)}$; \item
strong Taylor price, underlying $\hat{L}_{t}^{(1,\eps_1)}$; \item
frozen drift price, underlying $L_{t}^{(1,0)}$.
\end{itemize}
Numerical results in basis points (bps) are displayed in Table
\ref{table caplets} for parameters $\eps_1=1$, $N=3$,
$\a=0.50137$, $c_1=3.86777\%$, $c_2=3.7574\%$, $c_3=3.8631\%$,
$T_1=1.53151$, $T_i=T_1+i\a$, $i=2,3,4$. We characteristically
observe the difference in prices between the benchmark and frozen
drift price, whilst our strong Taylor correction method performs
very well and is computationally simpler and faster.
\begin{table}[ht]
{\small
\begin{tabular}{c|cccccc}
\emph{strikes}   & K=3\%  & K=3.5\% & K=4\% & K=5.75\% & K=6.25\% & K=8\%  \\
\hline\hline benchmark& 11.1831& 8.5897&  6.5503&  3.0349 & 2.4423& 1.2969 \\
\hline strong Taylor& 11.0687&  8.5691& 6.5867 & 3.1448 & 2.5513 & 1.3926\\
\hline frozen drift& 13.9551 & 11.1822 & 8.8803 & 4.6313 & 3.8506 &2.2524\\
\hline\hline
\end{tabular}}
\emph{\caption{Caplet values in bps for parameters $\eps_1=1$,
$\a=0.50137$, $c_1=3.86777\%$, $c_2=3.7574\%$, $c_3=3.8631\%$ and
$T_1=1.53151$.}}\label{table caplets}
\end{table}
\end{example}
%%%%%%%%%%%%%%%%%%%%%%%%%%%%%%%%%%%%%%%%%%%%%%%%%%%%%%%%%%%%%%%%%%%%%%%%%%%%%%%%%%%%%%%%%%%%%%%%%%%%%%%%%%%%%
\subsubsection{Weak Taylor Approximation} In what follows, we
provide some results on how to correct option prices obtained by
the SDE with the frozen drift \eqref{frozen sde} by adding a
correction term involving the appropriate Malliavin weight. Let
$\mathbf{L}_{T_i}^{i,k,\eps_1}$ denote the vector of the LIBOR
rates $(L_{T_i}^{(i,\eps_1)},\ldots,L_{T_i}^{(k,\eps_1)})$.
\begin{proposition}
Assume the setup of Lemma \ref{lemma freeze drift}, where the
$i^{th}$ LIBOR rate is given by:
\begin{equation}\label{real sde}
dL_{t}^{(i,\eps_1)}=
\sigma^{i}(t)L_{t}^{(i,\eps_1)}\Big(-\sum_{j=i+1}^{N} {\frac{\a
X_{t}^{(j,\eps_1)}\sigma^{j}(t)} {1+\a
X_{t}^{(j,\eps_1)}}}\rho_{ij}dt+dW_{t}^{i}\Big).
\end{equation}
with $L_0^{(i,\eps_1)}=X_0^{(i,\eps_1)}$ for all $\eps_1$ and all
$i=1,...,N$. Assume furthermore that the Malliavin covariance
matrix $\gamma(\mathbf{L}_{T_i}^{i,k,0})$ is invertible. Then the
price of an option with payoff $g(\mathbf{L}_{T_i}^{i,k,\eps_1})$,
for $i\leq k\leq N$ and $g$ bounded measurable, can be
approximated by the weak Taylor approximation of order one:
\begin{align}\label{multi price}
&\mathbf{W}_a^1(g,\mathbf{L}_{T_i}^{i,k,\eps_1})=
P(0,T)\Big(\e_{\mathbb{P}}\big(g(\mathbf{L}_{T_i}^{i,k,0})\big)
+\eps_1\e_{\mathbb{P}}\big(g(\mathbf{L}_{T_i}^{i,k,0})\zeta_{T_i}\big)\Big),
\end{align}
where the Malliavin weight $\zeta_{T_i}$ is given by:
\begin{align}\label{eq freeze multi
weights}
\zeta_{T_i}&=\delta\Big((D_{t}\mathbf{L}_{T_i}^{i,k,0})^{\mathrm{T}}
\gamma^{-1}(\mathbf{L}_{T_i}^{i,k,0})\frac{\partial}{\partial
\eps_1}\Big|_{\eps_1=0}\mathbf{L}_{T_i}^{i,k,\eps_1}\Big),
\end{align}
for $t\leq T_i$.

\begin{proof}
The weight $\zeta_{T_i}$ is obtained by \eqref{eq pi1}. Notice
that we can write:
\begin{equation*}
\frac{\partial}{\partial\eps_1
}\Big|_{\eps_1=0}\e_{\mathbb{P}}\big(g(\mathbf{L}_{T_i}^{i,k,\eps_1})\big)
=\e_{\mathbb{P}}\big(g\big(\mathbf{L}_{T_i}^{i,k,0}\big)\zeta_{T_i}\big),
\end{equation*}
and hence the result \eqref{multi price} by Definition \ref{def
weak taylor} for $n=1$.
\end{proof}\end{proposition}

%%%%%%%%%% example %%%%%%%%%%%%%%%%%%%%%%%%%%%%%%%%%%%%%%%%%%%%%%%%%%%%%
\begin{example}\label{eg freeze drift N=3}
In this example we let $N=3$ and we price a payers swaption with
strike price $K$ and maturity $T_1$, where the underlying swap is
entered at $T_1$ and has payment dates $T_2$ and $T_3$. We assume
that the volatility functions
$\sigma^i(t):[0,T_{i}]\rightarrow\mathbb{R}$ for $i=1,2,3$ are
constant:
\begin{equation*}
\sigma^1(t)=\sigma_1, \sigma^2(t)=\sigma_2, \sigma^3(t)=\sigma_3,
\end{equation*}
such that we obtain under the terminal measure $\mathbb{P}$:
\begin{align}\label{e-model}
dL_{t}^{(1,\eps_1)}&=\sigma_1 L_{t}^{(1,\eps_1)}\Big(
\rho_{12}\Big(-{\frac{\a X_{t}^{(2,\eps_1)}\sigma_2} {1+\a
X_{t}^{(2,\eps_1)}}}-{\frac{\a X_{t}^{(3,\eps_1)}\sigma_3} {1+\a
X_{t}^{(3,\eps_1)}}}\Big)dt+dW_{t}^{1}\Big),\nonumber\\
dL_{t}^{(2,\eps_1)}&=\sigma_2 L_{t}^{(2,\eps_1)}\Big(-\frac{\a
X_{t}^{(3,\eps_1)}\sigma_3} {1+\a
X_{t}^{(3,\eps_1)}}dt+dW_{t}^{2}\Big),\nonumber\\
&dL_{t}^{3}=\sigma_3 L_{t}^{3}dW_{t}^{2},\\
dX_{t}^{(1,\eps_1)}&=\eps_1\Big(\sigma_1
X_{t}^{(1,\eps_1)}\Big(\rho_{12}\Big(-{\frac{\a
X_{t}^{(2,\eps_1)}\sigma_2} {1+\a X_{t}^{(2,\eps_1)}}}-{\frac{\a
X_{t}^{(3,\eps_1)}\sigma_3} {1+\a
X_{t}^{(3,\eps_1)}}}\Big)dt+dW_{t}^{1}\Big)\Big),\nonumber\\
dX_{t}^{(2,\eps_1)}&=\eps_1\Big(\sigma_2
X_{t}^{(2,\eps_1)}\Big(-{\frac{\a X_{t}^{(3,\eps_1)}\sigma_3} {1+\a
X_{t}^{(3,\eps_1)}}}dt+dW_{t}^{2}\Big)\Big),\nonumber\\
dX_{t}^{(3,\eps_1)}&=\eps_1\Big(\sigma_3
X_{t}^{(3,\eps_1)}dW_{t}^{2}\Big)\nonumber,
\end{align}
with initial values $L_0^{(i,\eps_1)}=X_0^{(i,\eps_1)}=c_i$, for
$i=1,2,3$ and for all $\eps_1$. $W_{t}^{1}$ and $W_{t}^{2}$ are
correlated with correlation coefficient $\rho_{12}$. We freeze the
drifts in the above equations to obtain:
\begin{align*}
L_{t}^{(1,0)}&=c_1\exp{\Big(\sigma_1
W_{t}^{1}-\Big(\rho_{12}\big(\frac{\a c_2\sigma_2}{1+\a
c_2}+\frac{\a c_3\sigma_3}{1+\a
c_3}\big)+\frac{1}{2}\sigma_1\Big)\sigma_1t\Big)},\\
L_{t}^{(2,0)}&=c_2\exp{\Big(\sigma_2 W_{t}^{2}-\Big(\frac{\a
c_3\sigma_3}{1+\a c_3}+\frac{1}{2}\sigma_2\Big)\sigma_2t\Big)},\\
L_{t}^{3}&=c_3\exp{\Big(\sigma_3
W_{t}^{2}-\frac{1}{2}\sigma_3^2t\Big)}.
\end{align*}

Similarly to the previous example, we compare four option prices:
\begin{itemize}
\item benchmark price; \item frozen drift; \item strong Taylor
price; \item weak Taylor price.
\end{itemize}
The weak correction formula \eqref{multi price} adds a correction
term to the closed form price of the option. The swaption payoff
at $T_i$ can be found for example in \cite{musielarutkowski}:
\begin{equation*}
P_{T_i}^{swptn}=\Big(1-\sum_{k=i+1}^{N+1}\a_k
\prod_{j=i}^{k-1}(1+\a L_{T_i}^j)^{-1}\Big)_+,
\end{equation*}
if the underlying swap is entered at time $T_i$ and has payment
dates $T_{i+1},...,$ $T$. $\a_k$ is given by:
\begin{equation*}
\a_k= \left\{
\begin{array}{lll} K\a, & k=i+1,\ldots,N,\\1+K\a,  & k=N+1.
\end{array}\right.
\end{equation*}
The payers swaption value at time $t=0$ can be written as:
\begin{align}\label{general swaption price}
P&_{0}^{swptn}=P(0,T_{i})\e_{\mathbb{P}^{{i}}}\big(P_{T_i}^{swptn}\big)=\nonumber\\
&=P(0,T)\e_{\mathbb{P}}\Big(\Big(-\sum_{k=i}^{N}\a_k
\prod_{j=k}^{N}(1+\a L_{T_i}^j)-(1+K\a)\Big)_+\Big),
\end{align}
where $\a_i:=-1$ and $\mathbb{P}^{{i}}$ denotes the forward
measure corresponding to the bond $P(t,T_i)$ as num\'{e}raire.
Therefore, its benchmark price is given by the above formula with
$N=2$ and $i=1$:
\begin{equation*}
bP_0^{swptn}=P(0,T)\Bigg(\e_{\mathbb{P}}\Big(\Big(\a
L_{T_1}^{1}+\a L_{T_1}^{2} +\a^2L_{T_1}^{1}L_{T_1}^{2} -K\a^2
L_{T_1}^2-2K\a\Big)_+\Big)\Bigg).
\end{equation*}
Its weak Taylor price is given by \eqref{multi price} with $i=1$
and $k=N=2$.
\begin{align*}
wP_0^{swptn}&=P(0,T)\Bigg(\e_{\mathbb{P}}\Big(\Big(\a
L_{T_1}^{(1,0)}+\a L_{T_1}^{(2,0)}
+\a^2L_{T_1}^{(1,0)}L_{T_1}^{(2,0)} -K\a^2 L_{T_1}^{(2,0)}
-2K\a\Big)_+\Big)+\\
&+\eps_1\e_{\mathbb{P}}\Big(\Big(\a L_{T_1}^{(1,0)}+\a
L_{T_1}^{(2,0)} +\a^2L_{T_1}^{(1,0)}L_{T_1}^{(2,0)} -K\a^2
L_{T_1}^{(2,0)} -2K\a\Big)_+\zeta_{T_1}\Big)\Bigg).
\end{align*}
The weight $\zeta_{T_1}$ is given by \eqref{eq freeze multi
weights}. The partial derivative terms
$C_{T_1}^1:=\frac{\partial}{\partial
\eps_1}|_{\eps_1=0}L_{T_1}^{(1,\eps_1)}$ and
$C_{T_1}^2:=\frac{\partial}{\partial
\eps_1}|_{\eps_1=0}L_{T_1}^{(2,\eps_1)}$ are given by:
\begin{align*}
C_{T_1}^1&=L_{T_1}^{(1,0)}\int_{0}^{T_1}\sigma_1\rho_{12}\Big(\frac{\sigma_3\a
c_3\beta_2}{(1+\a c_3)}t -(\beta_2+\beta_3)W_{t}^{2}\Big)dt,
\end{align*}
and:
\begin{align*}
C_{T_1}^2&=L_{T_1}^{(2,0)}\int_{0}^{T_1}- \sigma_2 \beta_3
W_{t}^{2}dt,
\end{align*}
with $C_0^1=C_0^2=0$ and $\beta_2:=\frac{\a c_2^2\sigma_2^2}{(1+\a
c_2)^2}$, $\beta_3:=\frac{\a c_3^2\sigma_3^2}{(1+\a c_3)^2}$. The
Malliavin covariance matrix of the vector
$(L_{T_1}^{(1,0)},L_{T_1}^{(2,0)})$ is equal to:
\begin{align*}
&\gamma\big((L_{T_1}^{(1,0)},L_{T_1}^{(2,0)})\big)=\begin{pmatrix}
(1+\rho_{12}^2)(L_{T_1}^{(1,0)})^2T_1\sigma_1^2  &
2\rho_{12}(L_{T_1}^{(1,0)})(L_{T_1}^{(2,0)})T_1\sigma_1\sigma_2
\\ \\  2\rho_{12}(L_{T_1}^{(1,0)})(L_{T_1}^{(2,0)})T_1\sigma_1\sigma_2
& (1+\rho_{12}^2)(L_{T_1}^{(2,0)})^2T_1\sigma_2^2 \end{pmatrix}\\
&\Rightarrow
\textrm{det}\Big(\gamma\big((L_{T_1}^{(1,0)},L_{T_1}^{(2,0)})\big)\Big)
=(L_{T_1}^{(1,0)})^2(L_{T_1}^{(2,0)})^2T_1^2\sigma_1^2\sigma_2^2(1-\rho_{12}^2).
\end{align*}
The determinant is not zero as long as $\rho_{12}\neq 1$, which is a
natural assumption. Hence under this condition, its inverse is given
by:
\begin{align*}
&\gamma^{-1}\big((L_{T_1}^{(1,0)},L_{T_1}^{(2,0)})\big)=\frac{1}{(1-\rho_{12}^2)}\begin{pmatrix}
\frac{1+\rho_{12}^2}{(L_{T_1}^{(1,0)})^2T_1\sigma_1^2}&-\frac{2\rho_{12}}{L_{T_1}^{(1,0)}L_{T_1}^{(2,0)}T_1
\sigma_1\sigma_2}
\\ \\-\frac{2\rho_{12}}{L_{T_1}^{(1,0)}L_{T_1}^{(2,0)}T_1\sigma_1\sigma_2}
&
\frac{1+\rho_{12}^2}{(L_{T_1}^{(2,0)})^2T_1\sigma_2^2}\end{pmatrix}.
\end{align*}
Write the weight $\zeta_{T_1}=\zeta^1_{T_1}+\zeta^2_{T_1}$, where
the first weight $\zeta^1_{T_1}$ is obtained as:
\begin{equation*}
\zeta_{T_1}^1
=\int_{0}^{T_1}\Big(D_t^1L_{T_1}^{(1,0)}\big(C_{T_1}^1\gamma^{-1}_{11}
+C_{T_1}^2\gamma^{-1}_{12}\big)+D_t^1L_{T_1}^{(2,0)}\big(C_{T_1}^1\gamma^{-1}_{21}
+C_{T_1}^2\gamma^{-1}_{22}\big)\Big)\delta W_{t}^{1},
\end{equation*}
and $\zeta_{T_1}^2$ similarly:
\begin{equation*}
\zeta_{T_1}^2=
\int_{0}^{T_1}\Big(D_t^2L_{T_1}^{(1,0)}\big(C_{T_1}^1\gamma^{-1}_{11}
+C_{T_1}^2\gamma^{-1}_{12}\big)+D_t^2L_{T_1}^{(2,0)}\big(C_{T_1}^1\gamma^{-1}_{21}
+C_{T_1}^2\gamma^{-1}_{22}\big)\Big)\delta W_{t}^{2}.
\end{equation*}
Performing all necessary calculations, we conclude that:
\begin{align*}
\zeta_{T_1}^1&=\rho_{12}\Big(W_{T_1}^{1}\Big(\frac{\sigma_3\a
c_3\beta_2T_1}{2(1+\a
c_3)}-\frac{(\beta_2+\beta_3)}{T_1}\int_{0}^{T_1}W_{t}^{2}dt\Big)
+\frac{\rho_{12}(\beta_2+\beta_3)T_1}{2}\Big)-
\\&-\rho_{12}\Big(\frac{\rho_{12}\beta_3T_1}{2}-\frac{\beta_3
W_{T_1}^{1}}{T_1}\int_{0}^{T_1}W_{t}^{2}dt\Big).
\end{align*}
Analogously we obtain $\zeta_{T_1}^2$ as:
\begin{align*}
\zeta_{T_1}^2&=\rho_{12}^2\Bigg(W_{T_1}^{2}\Big(\frac{\sigma_3\a
c_3\beta_2T_1}{2(1+\a
c_3)}-\frac{(\beta_2+\beta_3)}{T_1}\int_{0}^{T_1}W_{t}^{2}dt\Big)
+\frac{(\beta_2+\beta_3)T_1}{2}\Bigg)-\\
&-\Big(\frac{\beta_3T_1}{2}-\frac{\beta_3
W_{T_1}^{2}}{T_1}\int_{0}^{T_1}W_{t}^{2}dt\Big).
\end{align*}
Notice that the weights are functions of normal variables and thus
the calculation of the weak Taylor price amounts just to
computation of deterministic integrals. Table \ref{table
swaptions3} gives the swaption prices in bps for parameters $N=3$,
$\a=0.25$, $\sigma_1=18\%$, $\sigma_2=15\%$, $\sigma_3=12\%$,
$c_0=5.28875\%$, $c_1=5.37375\%$, $c_2=5.40\%$, $c_3=5.40125\%$
and $\rho_{12}=0.75$.
\begin{table}[h]
{\small
\begin{tabular}{c|cccccc}
\emph{strikes}   & K=4\%  & K=4.5\% & K=4.75\% & K=5\% & K=5.15\% & K=5.25\%  \\
\hline\hline benchmark& 10.2240& 6.5386&  4.7454&  3.1060 & 2.2599& 1.7758 \\
\hline frozen drift &10.2132 & 6.5326 & 4.7419& 3.1028& 2.2582  & 1.7618\\
\hline strong Taylor& 10.2240& 6.5386&  4.7454&  3.1060 & 2.2599& 1.7758 \\
\hline weak Taylor& 10.2266 & 6.5407& 4.7485 & 3.1064 & 2.2593 & 1.7626\\
\hline\hline
\end{tabular}}
\emph{\caption{Swaption values in bps for parameters $\eps_1=1$,
$\a=0.25$, $\sigma_1=18\%$, $\sigma_2=15\%$, $\sigma_3=12\%$,
$c_0=5.28875\%$, $c_1=5.37375\%$, $c_2=5.40\%$, $c_3=5.40125\%$
and $\rho_{12}=0.75$.}}\label{table swaptions3}
\end{table}
\end{example}
%%%%%%%%%%%%%%%%%%%%%%%%%%%%%%%%%%%%%%%%%%%%%%%%%%%%%%%%%%%%%%%%%%%%%%%%%%%%%%%%%%%%%%%%%%%%%%%%%%%%%%%%%%%%%
\subsection{The Stochastic Volatility LIBOR Market
Model}\label{section SVLMM} In this section, we develop a
stochastic volatility LMM. The stochastic volatility parameter
$v_{t}$ follows a square root process, like in the extensively
applied Heston model \cite{heston93}. The resulting model, called
hereafter the \emph{stochastic volatility LMM} (SVLMM), has the
following dynamics under the terminal measure:
\begin{align}\label{sv terminal model}
dL_{t}^{i}&=
\sigma^{i}(t)L_{t}^{i}\sqrt{v_{t}}\Big(-\sum_{j=i+1}^{N}{\frac{\a
L_{t}^{j}\sigma^{j}(t)} {1+\a
L_{t}^{j}}}\rho_{ij}\sqrt{v_{t}}dt+dW_{t}^{i}\Big),i=1,...,N,\\
dv_{t}&=\kappa(\theta-v_{t})dt+\eps_2 \sqrt{v_{t}}dB_{t},\nonumber
\end{align}
where $\kappa,\theta,\eps_2\in\mathbb{R}_+$. The Brownian motions
$W_{t}=(W_t^{1},...,W_t^{N})$ and $B_{t}$ are expressed under the
terminal measure with correlations $dW^{i}_{t}dB_{t}=\rho_{i}dt$
and $dW^{i}_{t}dW_{t}^{j}=\rho_{ij}dt$ for $i,j=1,...N$. We assume
additionally that the filtration $(\mathcal{F}_t)_{0\leq t\leq T}$
is generated by both Brownian motions. Observe that the process
$v_t$ is a time-changed squared Bessel process with dimension
$\delta=4\kappa\theta/\eps_2^2$. If $\delta\geq 2$, then the point
zero is unattainable. So we require $2\kappa\theta\geq\eps_2^2$
for the process $v_{t}$ not to reach zero.
%%%%%%%%%%%%%%%%%%%%%%%%%%%%%%%%%%%%%%%%%%%%%%%%%%%%%%%%%%%%%%%%%%%%%%%%%%%%%%%%%%
\subsubsection{Pricing a multi-LIBOR option}\label{section multi
libor pricing} In this section, we aim at approximating the price
of an option with payoff depending on the vector
$\mathbf{L}_{T_i}^{i,k,\eps_1,\eps_2}=(L_{T_i}^{i,\eps_1,\eps_2},
\ldots,L_{T_i}^{k,\eps_1,\eps_2})$. We interpret the volatility of
the volatility parameter $\eps_2$ as a parameter on which the
LIBOR rates depend. Overall, we parametrise the SVLMM by both
$\eps_1$ and $\eps_2$ and correct prices in a weak sense
introducing Malliavin weights.

\begin{proposition}\label{prop multi-libor pricing}
Consider the SVLMM \eqref{sv terminal model} and assume that the
Malliavin covariance matrix $\gamma(\mathbf{L}_{T_i}^{i,k,0,0})$
is invertible. Then the price of an option with payoff
$\psi(\mathbf{L}_{T_i}^{i,k,\eps_1,\eps_2})$, $i\leq k\leq N$,
where $\psi$ is a bounded measurable function, can be approximated
by the weak Taylor approximation of order one:
\begin{align}\label{multi-libor pricing formula}
\mathbf{W}^1_{(\eps_1,\eps_2)}(\psi,\mathbf{L}_{T_i}^{i,k,\eps_1,\eps_2}))&=
P(0,T)\Big(\e_{\mathbb{P}}\big(\psi(\mathbf{L}_{T_i}^{i,k,0,0})\big)+
\eps_1\e_{\mathbb{P}}\big(\psi(\mathbf{L}_{T_i}^{i,k,0,0})\zeta_{T_i}\big)+\nonumber\\&+
\eps_2\e_{\mathbb{P}}\big(\psi(\mathbf{L}_{T_i}^{i,k,0,0})\pi_{T_i}\big)\Big),
\end{align}
where the Malliavin weights $\zeta_{T_i},\pi_{T_i}$ are given by:
\begin{align}\label{eq malliavin weight sv zeta i multi}
\zeta_{T_i}&=\delta\Big((D_{t}\mathbf{L}_{T_i}^{i,k,0,0})^{\mathrm{T}}
\gamma^{-1}(\mathbf{L}_{T_i}^{i,k,0,0}) \frac{\partial}{\partial
\eps_1}\Big|_{\eps_1=0}\mathbf{L}_{T_i}^{i,k,\eps_1,0}\Big),
\\ \label{eq malliavin weight sv pi i multi}
\pi_{T_i}&=\delta\Big((D_{t}\mathbf{L}_{T_i}^{i,k,0,0})^{\mathrm{T}}
\gamma^{-1}(\mathbf{L}_{T_i}^{i,k,0,0}) \frac{\partial}{\partial
\eps_2}\Big|_{\eps_2=0}\mathbf{L}_{T_i}^{i,k,0,\eps_2}),
\end{align}
for $t\leq T_i$.

\begin{proof}
The weights $\zeta_{T_i}$ and $\pi_{T_i}$ are obtained by
\eqref{eq pi1}. We derive \eqref{multi-libor pricing formula} by
noticing that:
\begin{align*}
&\e\big(\psi(\mathbf{L}_{T_i}^{i,k,\eps_1,\eps_2})\big)\simeq\\&=
\e\big(\psi(\mathbf{L}_{T_i}^{i,k,0,0})\big)
+\eps_1\frac{\partial}{\partial\eps_1
}\Big|_{\eps_1=0}\e\big(\psi(\mathbf{L}_{T_i}^{i,k,\eps_1,0})\big)+\eps_2\frac{\partial}{\partial\eps_2
}\Big|_{\eps_2=0}\mathbb{E}\big(\psi(\mathbf{L}_{T_i}^{i,k,0,\eps_2})\big)=\\
&=\e\big(\psi(\mathbf{L}_{T_i}^{i,k,0,0})\big)+
\eps_1\e\big(\psi(\mathbf{L}_{T_i}^{i,k,0,0})\zeta_{T_i}\big)+
\eps_2\e\big(\psi(\mathbf{L}_{T_i}^{i,k,0,0})\pi_{T_i}\big),
\end{align*}
from Definition \ref{def weak taylor} for $n=1$.
\end{proof}\end{proposition}
%%%%%%%%%%%%%%%%%%%%%%%% example %%%%%%%%%%%%%%%%%%%%%%%%%%%%%%%%%%%%%%%%%%%%%%%%%%%%%%%%%%%%%%%%%%%%%%%%%%%%%%%%%
\begin{example}
Let $N=2$ and consider the SVLMM where the volatility functions
$\sigma^i(t):[0,T_{i}]\rightarrow\mathbb{R}$ for $i=1,2$ are
assumed to be constant and in particular $ \sigma^1(t)=\sigma_1$,
$\sigma^2(t)=\sigma_2$. We derive an approximative formula for the
price of a payers swaption with maturity $T_1$ and strike price
$K$. The underlying swap is entered at $T_1$ and has payment dates
$T_2,T_3$. Under the terminal measure $\mathbb{P}$ we can write
the SDEs for the LIBOR rates and stochastic volatility as:
\begin{align*}
dv_{t}^{\eps_2}&=\kappa\Big(\theta-v_{t}^{\eps_2}\Big)dt+\eps_2
\sqrt{v_{t}^{\eps_2}}dB_{t},\\
dL_{t}^{(1,\eps_1,\eps_2)}&=-L_{t}^{(1,\eps_1,\eps_2)}\rho_{12}\frac{\a
X_{t}^{(2,\eps_1,\eps_2)}\sigma_2}{1+\a
X_{t}^{(2,\eps_1,\eps_2)}}\sigma_1v_{t}^{\eps_2}dt+\sigma_1
L_{t}^{(1,\eps_1,\eps_2)}\sqrt{v_t^{\eps_2}}dW_{t}^{1},\\
dL_{t}^{(2,\eps_2)}&=\sigma_2
L_{t}^{(2,\eps_2)}\sqrt{v_t^{\eps_2}}dW_{t}^{2},\\
dX_{t}^{(2,\eps_1)}&=\eps_1\Big(\sigma_2
X_{t}^{(2,\eps_2)}\sqrt{v_t^{\eps_2}}dW_{t}^{2}\Big).
\end{align*}
$W_{t}^{1}$ and $W_{t}^{2}$ are assumed to be correlated, so
correlations are as $dW_{t}^{i}dB_{t}=\rho_i dt$ and
$dW_{t}^{1}dW_{t}^{2}=\rho_{12}$ for $i=1,2$. The $(0,0)$-model is
given by:
\begin{align*}
v_t^{0}&=\exp{(-\kappa t)}(v_0^{0}-\theta)+\theta,\\
L_{T_1}^{(1,0,0)}&=
c_1\exp{\Big(\sigma_1\int_{0}^{T_{1}}\sqrt{v_t^{0}}
dW_t^{1}-\big(\frac{\a c_2\rho_{12}}{1+\a c_2}\sigma_2+\frac{1}{2}\sigma_1\big)c\sigma_1\Big)},\\
L_{T_1}^{(2,0)}
&=c_2\exp{\Big(\sigma_2\int_{0}^{T_{1}}\sqrt{v_t^{0}}
dW_t^{2}-\frac{1}{2}\sigma_2^2c\Big)},\\
X_t^{(2,0,0)}&=c_2,
\end{align*}
with $c:=\int_{0}^{T_{1}}v_t^{0}dt=\theta
T_1-\frac{v_0^{0}-\theta}{\kappa}(\exp{(-\kappa T_1)}-1)$. As in
the previous example, we compare the following option prices:
\begin{itemize}
\item benchmark price; \item frozen drift; \item weak Taylor price
\eqref{multi-libor pricing formula}.
\end{itemize}
The benchmark price is given by \eqref{general swaption price}
with $N=2$ and $i=1$:
\begin{equation*}
bP_0^{swptn}=P(0,T)\e_{\mathbb{P}}\Big(\big(\a
L_{T_1}^{(1,\eps_1,\eps_2)}+\a
L_{T_1}^{(2,\eps_2)}+\a^2L_{T_1}^{(1,\eps_1,\eps_2)}L_{T_1}^{(2,\eps_2)}
-K\a^2 L_{T_1}^{(2,\eps_2)}-2K\a\big)_+\Big).
\end{equation*}
The weak Taylor price is obtained by \eqref{multi-libor pricing
formula}:
\begin{align*}\label{eq temp swp}
wP_0^{swptn}&=P(0,T)\Bigg(\e_{\mathbb{P}}\Big(\big(\a
L_{T_1}^{(1,0,0)}+\a
L_{T_1}^{(2,0)}+\a^2L_{T_1}^{(1,0,0)}L_{T_1}^{(2,0)}-K\a^2
L_{T_1}^{(2,0)}-\nonumber\\
&-2K\a\big)_+\Big)+\eps_1\e_{\mathbb{P}}\Big(\big(\a
L_{T_1}^{(1,0,0)}+\a
L_{T_1}^{(2,0)}+\a^2L_{T_1}^{(1,0,0)}L_{T_1}^{(2,0)}-K\a^2\cdot\nonumber\\
&\cdot
L_{T_1}^{(2,0)}-2K\a\big)_+\zeta_{T_1}\Big)+\eps_2\e_{\mathbb{P}}\Big(\big(\a
L_{T_1}^{(1,0,0)}+\a
L_{T_1}^{(2,0)}+\a^2L_{T_1}^{(1,0,0)}L_{T_1}^{(2,0)}-\nonumber\\
&-K\a^2 L_{T_1}^{(2,0)}-2K\a\big)_+\pi_{T_1}\Big)\Bigg).
\end{align*}
We calculate the Malliavin weights $\zeta_{T_1}$, $\pi_{T_1}$ as
given by \eqref{eq malliavin weight sv zeta i multi} and \eqref{eq
malliavin weight sv pi i multi} correspondingly. We can express
the weight $\zeta_{T_1}$ as:
\begin{align*}
\zeta_{T_1}&=\zeta_{T_1}^1+\zeta_{T_1}^2,
\end{align*}
with:
\begin{align*}
\zeta_{T_1}^1&=\int_{0}^{T_1}\Bigg(\Big(D_t^1L_{T_1}^{(1,0,0)}\frac{\partial}{\partial\eps_1}\Big|_{\eps_1=0}L_{T_1}^{(1,\eps_1,0)}
\gamma^{-1}(L_{T_1}^{(1,0,0)},L_{T_1}^{(2,0,0)})_{11}\Big)+\\
&+\Big(D_t^1L_{T_1}^{(2,0,0)}\frac{\partial}{\partial\eps_1}\Big|_{\eps_1=0}L_{T_1}^{(1,\eps_1,0)}
\gamma^{-1}(L_{T_1}^{(1,0,0)},L_{T_1}^{(2,0,0)})_{21}\Big)\Bigg)\delta
W_{t}^1,
\end{align*}
and:
\begin{align*}
\zeta_{T_1}^2&=\int_{0}^{T_1}\Bigg(\Big(D_t^2L_{T_1}^{(1,0,0)}\frac{\partial}{\partial\eps_1}\Big|_{\eps_1=0}L_{T_1}^{(1,\eps_1,0)}
\gamma^{-1}(L_{T_1}^{(1,0,0)},L_{T_1}^{(2,0,0)})_{11}\Big)+\\
&+\Big(D_t^2L_{T_1}^{(2,0,0)}\frac{\partial}{\partial\eps_1}\Big|_{\eps_1=0}L_{T_1}^{(1,\eps_1,0)}
\gamma^{-1}(L_{T_1}^{(1,0,0)},L_{T_1}^{(2,0,0)})_{21}\Big)\Bigg)\delta
W_{t}^2.
\end{align*}
since $\frac{\partial}{\partial\eps_1}|_{\eps_1=0}$
$L_{T_1}^{(2,\eps_1,0)}=0$. The partial derivative term with
respect to $\eps_1$ for $L^1$ is given by:
\begin{align*}
&\frac{\partial}{\partial\eps_1}\Big|_{\eps_1=0}L_{T_1}^{(1,\eps_1,0)}=
L_{T_1}^{(1,0,0)}\int_{0}^{T_1}\sigma_1 v_{t}^{0}\Big(
-\frac{\a\sigma_2^2c_2^2\rho_{12}}{(1+\a
c_2)^2}\int_{0}^{t}\sqrt{v_{s}^{0}}dW_{s}^{2}\Big)dt=\\
&=-\sigma_1\rho_{12}\beta_2L_{T_1}^{(1,0,0)}
\int_{0}^{T_1}\sqrt{v_{t}^{0}}\Big(\theta(T_1-t)
-\frac{v_0^{0}-\theta}{\kappa}(\exp{(-\kappa T_1)}-\exp{(-\kappa
t)})\Big)dW_{t}^{2},
\end{align*}
where $\beta_2=\frac{\a\sigma_2^2c_2^2}{(1+\a c_2)^2}$. Similarly
the weight $\pi_{T_1}$ is given by:
\begin{align*}
\pi_{T_1}&=\pi_{T_1}^1+\pi_{T_1}^2,
\end{align*}
with:
\begin{equation*}
\pi_{T_1}^1=\int_{0}^{T_1}\big(\sum_{l=1}^{2}D_t^1L_{T_1}^{(l,0,0)}
\sum_{j=1}^{2}\frac{\partial}{\partial\eps_2}\Big|_{\eps_2=0}L_{T_1}^{(j,0,\eps_2)}
\gamma^{-1}\big((L_{T_1}^{(1,0,0)},L_{T_1}^{(2,0,0)})_{lj}\big)
\big)\delta W_{t}^1,
\end{equation*}
and:
\begin{equation*}
\pi_{T_1}^2=\int_{0}^{T_1}\big(\sum_{l=1}^{2}D_t^2L_{T_1}^{(l,0,0)}
\sum_{j=1}^{2}\frac{\partial}{\partial\eps_2}\Big|_{\eps_2=0}L_{T_1}^{(j,0,\eps_2)}
\gamma^{-1}\big((L_{T_1}^{(1,0,0)},L_{T_1}^{(2,0,0)})_{lj}\big)
\big)\delta W_t^2.
\end{equation*}

Partial derivative terms are equal to:
\begin{align*}
&\frac{\partial}{\partial\eps_2}\Big|_{\eps_2=0}L_{T_1}^{(1,0,\eps_2)}=
L_{T_1}^{(1,0,0)}\Big(\frac{\sigma_1}{2}
\int_{0}^{T_1}\frac{\exp{(-\kappa
t)}}{\sqrt{v_t^{0}}}\int_{0}^{t}\exp{(\kappa
s)}\sqrt{v_s^{0}}dB_sdW_t^{1}+\\&+
\frac{1}{\kappa}(\frac{\sigma_1^2}{2}+\frac{\a
c_2\sigma_1\sigma_2\rho_{12}}{1+\a c_2})\int_0^{T_1}\exp{(\kappa
s)}\sqrt{v_s^{0}}\big(\exp{(-\kappa T_1)}-\exp{(-\kappa
s)}\big)dB_s\Big).
\end{align*}
Doing similar calculations,
we derive the second partial derivative:
\begin{equation*}
\frac{\partial}{\partial\eps_2}\Big|_{\eps_2=0}L_{T_1}^{(2,\eps_2)}=L_{T_1}^{(2,0)}\Big(\frac{1}{2}
\int_{0}^{T_1}\frac{\sigma_2}{\sqrt{v_t^{0}}}V_tdW_t^{2}-
\frac{1}{2}\int_0^{T_1}\sigma_1\sigma_2V_tdt\Big),
\end{equation*}
where $V_t=\exp{(-\kappa t)}\int_{0}^{t}\exp{(\kappa
s)}\sqrt{v_s^{0}}dB_s$.

We calculate the Malliavin covariance matrix
$\gamma\big((L_{T_1}^{(1,0,0)},L_{T_1}^{(2,0,0)})\big)$
 and its inverse.
\begin{align*}
&\gamma=\begin{pmatrix}
(1+\rho_{12}^2)(L_{T_1}^{(1,0,0)})^2\sigma_1^2\underbrace{\int_{0}^{T_1}v_t^0dt}_{=c}
& 2\rho_{12}L_{T_1}^{(1,0,0)}L_{T_1}^{(2,0,0)}\sigma_1\sigma_2c
\\ \\2\rho_{12}L_{T_1}^{(1,0,0)}L_{T_1}^{(2,0,0)}\sigma_1\sigma_2c
& (1+\rho_{12}^2)(L_{T_1}^{(2,0,0)})^2\sigma_2^2c\end{pmatrix}\\
&\Rightarrow
\textrm{det}\Big(\gamma\big((L_{T_1}^{(1,0,0)},L_{T_1}^{(2,0,0)})\big)\Big)
=(L_{T_1}^{(1,0,0)})^2(L_{T_1}^{(2,0,0)})^2\sigma_1^2\sigma_2^2c^2(1-\rho_{12}^2).\nonumber
\end{align*}
Hence its inverse is given by, for $\rho_{12}\neq 1$:
\begin{align*}
&\gamma^{-1}=\frac{1}{(1-\rho_{12}^2)}\begin{pmatrix}
\frac{1+\rho_{12}^2}{(L_{T_1}^{(1,0,0)})^2\sigma_1^2c} &
-\frac{2\rho_{12}}{L_{T_1}^{(1,0,0)}L_{T_1}^{(2,0,0)}\sigma_1\sigma_2c}
\\ \\-\frac{2\rho_{12}}{L_{T_1}^{(1,0,0)}L_{T_1}^{(2,0,0)}\sigma_1\sigma_2c}
&
\frac{1+\rho_{12}^2}{(L_{T_1}^{(2,0,0)})^2\sigma_2^2c}\end{pmatrix}.
\end{align*}
If we define $X_i=\int_{0}^{T_1}\sqrt{v_{t}^{0}}dW_{t}^{i}$,
$i=1,2$ and $Y=\int_{0}^{T_1}\sqrt{v_{t}^{0}}\Big(\theta(T_1-t)
-\frac{v_0^{0}-\theta}{\kappa}(\exp{(-\kappa T_1)}-\exp{(-\kappa
t)})\Big)dW_{t}^{2}$, we finally obtain the weights as:
\begin{equation*}
\zeta_{T_1}^1=-\frac{\rho_{12}\beta_2}{c}\Big(X_1Y-\cov(X_1,Y)\Big),
\end{equation*}
and:
\begin{equation*}
\zeta_{T_1}^2=\frac{\rho_{12}^2\beta_2}{c}\Big(X_2Y-\cov(X_2,Y)\Big).
\end{equation*}
Moreover, for the weight $\pi_{T_1}$ we define:
\begin{equation*}
B=\int_0^{T_1}g(t)\Big(\exp{(-\kappa T_1)}-\exp{(-\kappa
t)}\Big)dB_t,
\end{equation*}
and random variables $D_i$, $Z_i$ for $i=1,2$:
\begin{align*}
D_i&=\int_{0}^{T_1}f(t)\int_{0}^{t}g(s)dW_s^{i}dW_t^{i},\\
Z_i&=\int_{0}^{T_1}f(t)\int_{0}^{t}g(s)dZ_s^{i}dW_t^{i},
\end{align*}
where the Brownian motions $Z_t^i$ are independent from $W_t^i$
and $f(t)=\frac{\exp{(-\kappa t)}}{\sqrt{v_t^{0}}}$,
$g(s)=\exp{(\kappa s)}\sqrt{v_s^{0}}$. Therefore, we obtain the
weights as:
\begin{align*}
\pi_{T_1}^1&=\frac{1}{2c}\Bigg(X_1(\rho_1D_1+\sqrt{1-\rho^2_1}Z_1)+
\frac{BX_1}{\kappa}\Big(\frac{\a
c_2(2\rho_{12}\sigma_2+\sigma_1)+\sigma_1}{1+\a
c_2}\Big)-\\
&-\frac{\rho_1E}{\kappa}\Big(\frac{\a
c_2(2\rho_{12}\sigma_2+\sigma_1)+\sigma_1}{1+\a
c_2}\Big)\Bigg)-\frac{\rho_{12}}{2c}\Bigg(X_1(\rho_2D_2+\sqrt{1-\rho^2_2}Z_2)+\\
&+\frac{\sigma_1
BX_1}{\kappa}+\frac{\rho_{12}B}{\kappa}-\frac{\sigma_1\rho_1E}{\kappa}\Bigg),
\end{align*}
where $E$ equals to $\Big(\frac{1}{\kappa}\big(1-\exp{(-\kappa
T_1)}\big)-T_1\Big)$. Similarly we get $\pi_{T_1}^{2}$ as:
\begin{align*}
\pi_{T_1}^{2}&=\frac{1}{2c}\Bigg(\frac{\sigma_2}{\sigma_1}X_2(\rho_2D_2+\sqrt{1-\rho^2_2}Z_2)+
\frac{B}{\kappa}\big(\sigma_2
X_2+1\big)-\frac{\sigma_2\rho_2E}{\kappa}\Bigg)+\\
&-\frac{\rho_{12}}{2c}\Bigg(\frac{\sigma_1}{\sigma_2}X_2(\rho_1D_1+\sqrt{1-\rho^2_1}Z_1)+
\frac{\sigma_1 BX_2}{\kappa\sigma_2}\Big(\frac{\a
c_2(2\rho_{12}\sigma_2+\sigma_1)+\sigma_1}{1+\a
c_2}\Big)+\\
&+\frac{\rho_{12}B}{\kappa}-
\frac{\sigma_1\rho_2E}{\kappa\sigma_2}\Big(\frac{\a
c_2(2\rho_{12}\sigma_2+\sigma_1)+\sigma_1}{1+\a c_2}\Big)\Bigg).
\end{align*}

In this example, the weights are functions of normal variables and
double stochastic integrals, which are computed via simulation.
Table \ref{table swaptionsv} reports the swaption prices in bps
with parameters $N=2$, $\a = 1.5$, $\sigma_1=25\%$,
$\sigma_2=15\%$, $c_0=5.28875\%$, $c_1=5.4\%$, $c_2=5.39\%$,
$v_0=1$, $\rho_1=-0.75$, $\rho_2=-0.6$, $\kappa=2.3767$,
$\theta=0.2143$, $\eps_2=25\%$, $\rho_{12}=0.63$.
\begin{table}[ht]
{\small \begin{tabular}{c|cccccc} \emph{strikes}  &K=3.5\%  & K=4\%  &K=5\%& K=6\% & K=7\% &K=8\% \\
\hline\hline benchmark & 3.8984& 2.9221&1.2588&0.3858& 0.1019 &0.0216\\
\hline $(0,0)$-model & 3.8951& 2.9053 &1.2705 & 0.3966 & 0.0942 & 0.0185\\
\hline weak Taylor & 3.8990& 2.9159& 1.2694& 0.3791 & 0.1042& 0.0210\\
\hline\hline
\end{tabular}}
\emph{\caption{Stochastic volatility swaption values in bps for
parameters $\eps_1 = 1$, $\a = 1.5$, $\sigma_1=25\%$,
$\sigma_2=15\%$, $c_0=5.28875\%$, $c_1=5.4\%$, $c_2=5.39\%$,
$v_0=1$, $\rho_1=-0.75$, $\rho_2=-0.6$, $\kappa=2.3767$,
$\theta=0.2143$, $\eps_2=25\%$, $\rho_{12}=0.63$.}}\label{table
swaptionsv}
\end{table}

\end{example}

%%%%%%%%%%%%%% bibliography %%%%%%%%%%%%%%%%%%%%%%%%%%%%%%%%%%%%%%%%%%%%%%%%%%%%%%%%%%%%%%%%%%%%%%%%%%%%%%%%%
\bibliographystyle{amsalpha}
\bibliography{rates}

\end{document}